\documentclass{ws-ijmpd}

\begin{document}

\markboth{J. A. de Freitas Pacheco, T. Regimbau, S. Vincent and A. Spallicci}
{Expected coalescence rates of NS-NS binaries for laser beam interferometers}

\catchline{}{}{}{}{}
%
%%%%%%%%%%%%%%%%%%%%%%%%%%%%%%%%%%%%%%%%%%%%%%%%%%%%%%%%%%%%%%%%%%%%

\title{Expected coalescence rates of NS-NS binaries for laser beam interferometers}
\author{J. A. de Freitas Pacheco$^1$, T. Regimbau$^{2,3}$,
  S. Vincent$^1$ and A. Spallicci$^2$}
\address{
$^1$UMR 6202 Cassiop\'ee, CNRS, Observatoire de la C\^ote d'Azur, BP 4229, 06304 NICE Cedex 4 (France)\\
$^2$UMR 6162 Artemis, CNRS, Observatoire de la C\^ote d'Azur, BP 4229, 06304 NICE Cedex 4 (France)\\
$^3$School of Physics and Astronomy, Cardiff University, 5 The Parade, Cardiff,
CF2 3YB, United Kingdom \\
pacheco@obs-nice.fr, regimbau@obs-nice.fr, vincent@obs-nice.fr, spallicci@obs-nice.fr}

\maketitle

\begin{history}
\received{Day Month Year}
\revised{Day Month Year}
\comby{Managing Editor}
\end{history}

\begin{abstract}
The coalescence rate of two neutron stars (NS) is revisited. 
For estimation of the number of bound NS-NS and the probability of their 
coalescence in a timescale $\tau$, the galactic
star formation history, directly derived from observations, and the evolution of 
massive stars are considered. 
The newly established galactic merging rate is
$(1.7\pm 1.0) \times 10^{-5}\, yr^{-1}$, while the local merging rate,
including the contribution of elliptical galaxies, is about a factor of 
two higher, $3.4 \times 10^{-5}\, yr^{-1}$.
Using the present data basis on galaxy distribution in the local universe and the expected
sensitivity of the first generation of
laser beam interferometers, we estimate
that one event should occur every 125 years for LIGO and
one event each 148 years for VIRGO. The situation is considerably improved for
advanced-LIGO since we predict that 6 events per year should be detected whereas
for a recently proposed VIRGO new configuration, the
event rate might increase up to 3 events every two years. 
\end{abstract}


\section{Introduction}

The merger of  two neutron stars (NS-NS), two black holes (BH-BH ) or a 
black hole and a neutron star (BH-NS) are among the most important sources
of gravitational waves, due to the huge energy released in the
process. In particular, the coalescence of NS-NS binaries may radiate
about 10$^{53}$ erg in the last seconds of their inspiral trajectory,
at frequencies up to 1.4-1.6 kHz, range covered by most of the
ground-based laser interferometers like VIRGO\cite{bra}, LIGO\cite{abr}, GEO\cite{hou} or TAMA\cite{kur}.
For understanding of the process dynamics, the prediction of its signal 
waveform, the estimate of its released power and the event rate,
theoretical studies on the coalescence of NS-NS systems have been
performed  in the past years\cite{fab,ros,dam}. The dynamics of the
inspiral and, in particular, the study of tidal effects in the very
last orbits, may put strong constraints on the equation of state of
dense matter. Concerning the coalescence rate of NS-NS binaries,
theoretical evaluations are always performed in two main steps: firstly, the merging rate 
in our Galaxy is estimated and then, assuming 
that this rate is typical, an estimate is performed for the volume of
the universe sampled by a given detector, using some adequate
scaling. The galactic merging rate has been estimated either by using
binary population synthesis models\cite{pz1,pz2,bel} or from the
statistics  of the observed NS-NS binaries\cite{phi,vdh,kal1,kim,kal2}. These
estimates may differ by one order of magnitude, ranging in general
from 10$^{-6}$ yr$^{-1}$ up to few times 10$^{-5}$ yr$^{-1}$, but
values as high as $\sim 3\times 10^{-4}\, yr^{-1}$ have been reported in the literature\cite{tut,lip} .
Recently Ref.~\refcite{kal3} included the new highly relativistic
NS-NS binary system PSR J0737-3039 in their statistical analysis, obtaining a 
value ($\sim 8.3\times 10^{-5}\, yr^{-1}$) comparable to the high
rates derived previously by Ref.~\refcite{tut} and Ref.~\refcite{lip}. 
Once the present galactic merging rate  has been evaluated, the
expected detection rate can be estimated by scaling the total
luminosity within the volume probed by the detector with respect to
the luminosity of the Galaxy at the same wave-band. Within the local 
universe (z $<$ 0.01), the distribution of galaxies is not homogeneous and if the actual
distribution is not taken into account, the expected detection rate
may be seriously underestimated\cite{nut}. 
For instance, the inclusion of the huge concentration of galaxies in the direction of 
Norma-Centaurus may considerably increase the detection rate of these
events\cite{dfp}. Estimates of the galactic merging
rate performed according to a given evolutionary model, require the
knowledge of the star formation history. These models assume usually
that the star formation rate is proportional to the available mass of
gas and a consequent exponential decay with time. However,
observations and numerical simulations suggest that  "disky" galaxies,
including the Milky Way, are gradually formed by accretion and merger
episodes\cite{hel,pei}, which affect the star formation history.
It is clear that in spite of all efforts  made in the past years to
evaluate the local coalescence rate and the expected detection rate by
a given interferometric detector, the uncertainties are still quite
large with consequences  bearing on the future development of instruments and the 
detection strategy. In  the present work, we combine the advantages 
of different approaches to present a new estimate of  galactic merging rate 
based on population synthesis of the pulsar
population\cite{reg1,reg2} and on  new simulations on the evolution
of massive binaries. From the latter, we have derived the distribution
probability $P(\tau)$, which gives the fraction of newly formed  NS-NS
binaries with a coalescence timescale $\tau$, due to gravitational
radiation losses. For the first time the star formation history  of
our Galaxy, derived directly from observations, was included in the
computation of the galactic NS-NS merging rate. Since elliptical
galaxies contribute also to the morphological composition of galaxies
in the local universe, we have estimated  the expected coalescence
rate in these objects, by adopting a star formation model able to reproduce their 
observed photometric properties. Then, the local average coalescence rate
has been estimated and weighted according to the total light fraction
contribution of each morphological type. In a second step, we evaluate
the volume of the universe probed by a given detector and compute the
total luminosity inside this volume, using available data basis. The
organization of this paper is the following: in  Section II we present
the derivation of the local coalescence rate, in  Section  III we
estimate the expected detection rates taking into account the planned
sensibility of the different interferometers and, finally, in  Section IV we 
discuss our results and summarize our main conclusions.

\section{The local coalescence rate}
\subsection{The galactic rate}
 
In order to calculate the coalescence rate of NS-NS pairs in a given galaxy, we
adopt here the approach by Ref.~\refcite{dfp}. Let us suppose that a
massive binary (masses of components higher than 9 M$_{\odot}$) is formed at 
the instant $t'$. Let $\tau_*$  be the mean evolutionary 
timescale required for the system to evolve into two neutron stars,
typically of the order of $10^7 - 10^8$ yr. Define $P(\tau)$ as the
probability per unit of time for a newly formed NS-NS binary  to
coalesce in a timescale $\tau$ and define $R_*(t)$ as the star
formation rate, given in $M_{\odot}$.yr$^{-1}$. Under these
conditions, the coalescence rate at instant t is
\begin{equation}
\nu_c(t) = f_b\beta_{ns}\lambda\int_{\tau_0}^{(t-\tau_*-\tau_0)}P(\tau)R_*(t-\tau_*-\tau)d\tau 
\label{mergerate}
\end{equation}
where $f_b$ is the fraction of massive binaries formed among all
stars, $\beta_{ns}$ is the fraction of formed binaries which remain
bounded after the second supernova event and $\lambda$ is the fraction per unit mass
of formed stars in the mass interval 9-40 M$_{\odot}$. We assume that 
progenitors with initial masses above 40 M$_{\odot}$ will produce
black holes. We have set $\tau_0$ as the minimum timescale for a NS-NS
binary to coalesce. This minimum timescale was estimated from
simulations to be described below. If the initial mass function (IMF) is of the 
form $\xi(m) = Am^{-\gamma}$, with $\gamma \approx$ 2.35 \, (Salpeter's law),
normalized within the mass interval 0.1 - 80 M$_{\odot}$ such as $\int
m\xi(m)dm$ = 1, then $\lambda = \int_9^40\xi(m)dm = 5.72~\times 10^{-3}~ M_{\odot}^{-1}$.
The parameters $\beta_{ns}$, $\tau_0$ and the probability distribution
$P(\tau)$ were calculated from numerical simulations described in
detail by Ref.~\refcite{vin}. Here we outline only the main aspects of these simulations.
A massive binary is initially generated according to the following
prescriptions. The mass of the primary is obtained from a probability
distribution corresponding to a Salpeter's IMF, while the secondary
mass is derived from the observed mass ratio distribution for massive
binaries obtained from observations~\cite{tri}. The initial pair
separation {\it a} is fixed by a probability distribution $P(a)
\propto da/a$~~\cite{kra}, normalized between R$_{min}$ and R$_{max}$. The minimum separation 
was taken to be equal to twice the Roche lobe
of the primary and the maximum separation was taken to be equal to
100R$_{min}$. The initial eccentricity of the orbit was assumed to
obey a distribution probability $P(e)de = 2ede$, corresponding to
orbits filling completely the phase space. Until the explosion of the
first supernova (the primary star), the orbital parameters of the
system vary due to the mass-loss produced by the stellar wind from
both stars or mass transfer. The latter mechanism is more rare, since
these massive stars do not frequently reach or overlap the
Roche-lobe. The explosion of the first supernova leaves a NS remnant
of 1.4 M$_{\odot}$ and produces an "instantaneous" mass loss, which
may disrupt (total orbital energy positive after the event) or not
(total orbital energy negative after the explosion) the originally
bound pair. Our simulations indicate that 77.4\% of the systems remain
bounded after the first supernova. The bound binary, now constituted
by a NS star and an evolved star, which has already lost part of its
original mass, enters into a second phase of slow mass-loss, with the orbital parameters 
varying as before, until the second supernova explosion occurs. We recall that
our evolutionary scenario is similar to that developed by Ref.~\refcite{bel},
in  which none of the stars ever had the chance of being recycled by accretion. 
We have performed 500,000 numerical experiments, from which 
resulted 11,627 NS-NS bound pairs, implying  $\beta_{ns}$ = 0.024.
A velocity kick is probably imparted to the nascent neutron star
although its mechanism is still a matter of debate. Natal kicks may
unbind binaries which otherwise might have remained bound or, less
probably, conserve bound systems which without the kick would have been disrupted. In these 
experiments we have assumed that neutron stars have
a natal kick velocity corresponding to a 1-D velocity dispersion of
about 80 km/s. We have also investigated how these results are changed if a Maxwellian
velocity distribution with a dispersion 1-D of 230 km/s is adopted. In this case, the resulting 
fraction of bound binaries is reduced by one order of magnitude, e.g., $\beta_{ns}$ =
0.0029. These numbers corresponding to the resulting fraction of bound systems are consistent with 
previous analyses on effects of the natal kick~\cite{lip,pz2}. Recent investigations on the
spin period-eccentricity relation for NS-NS systems~\cite{dpp05} suggest that such a
correlation can only be obtained if the second neutron star receives a kick substantially
smaller (velocity dispersion less than 50 km/s) than kick velocities commonly assumed for single
radio pulsars. If this conclusion is correct, values of $\beta_{ns}$ derived from
simulations with low kick velocity dispersion are more likely and will be adopted here.
Clearly, the natal kick amplitude remains the major source of uncertainty in the estimate of 
the fraction of bound NS-NS binaries. 
However, once the relative number of NS-NS binaries is fixed, the
fraction $f_b$ of massive binaries formed among all stars is related to $\beta_{ns}$ through the equation 
\begin{equation}
\frac{N_p}{N_b} = \frac{1}{\beta_{ns}}\frac{(1-f_b)}{f_b} + 2\frac{(1-\beta_{ns})}
{\beta_{ns}}
\end{equation}
where $N_p$ and $N_b$ are respectively the number of single pulsars and the number of NS-NS 
binaries in the Galaxy. We have taken into account that the population of single pulsars 
results not only from the evolution of single stars but also partially
from disrupted massive binaries. Since $f_b$ = 1 is a strong upper
limit, this imposes a lower limit for $\beta_{ns}$ for a given ratio
$N_p/N_b$. The number of isolated pulsars N$_p$ has been derived from
population synthesis of single radio pulsars. We have employed the code described in detail 
in Ref.~\refcite{reg1,reg2}, up-graded to take into account the new pulsars discovered at high frequencies 
by the Parkes Multibeam Survey as well as the most recent model for the pulsar velocity 
distribution~\cite{arz}. The method consists in generating population of pulsars with a given
set of birth properties, following their evolution according to the
dipole braking model and modeling selection effects that constrain
radio detection. The best agreement between simulated and observed
distributions of physical properties as the period, period derivative,
distance, among others fixes the initial period and magnetic braking timescale distributions 
(see parameters in Table 1), as well as the total number of pulsars in the 
Galaxy ($N_p \approx 250000$) and their birthrate (one pulsar every 90 yr).
Simulations using the recent results by Ref.~\refcite{hobbs} for the
velocity distribution of radio pulsars were also performed and no significant differences in
the values derived for parameters shown in Table 1 or for the total number of pulsars were
noticed.
\begin{table}[ph]
\tbl{parameters of the initial period $P_0$ and magnetic braking timescale 
ln $\tau_0$ distribution, assumed to be Gaussian}
{\begin{tabular}{@{}ccc@{}} \toprule
mean  &  dispersion  \\ \colrule
$P_0 (ms) = 240 \pm 20$&$\sigma_{P_0} = 80 \pm 20$\\
$ln \tau_0 (s) = 11 \pm 0.5$&$\sigma_{ln\tau_0} = 3.6 \pm 0.2$\\\botrule
\end{tabular} \label{ta1}}
\end{table}

The number of NS-NS binaries can be derived by linking the properties of 
the youngest pulsar in a given double neutron star system, supposed
not to have been affected by external torques, to those of the
population of single pulsars. In other words, if the evolution of the
newly formed (second) pulsar was not yet affected by torques other
than the canonical magnetic dipole, its rotation period increases as those of isolated pulsars. 
For this set of simulations, we have adopted a velocity distribution for the barycenter
of the NS-NS binaries different from the velocity distribution adopted for single pulsars.
We consider that millisecond pulsars (MSP) were spun-up by mass transfer, before the second
supernova explosion, which disrupted the system. According to Ref.~\refcite{lyn1}
MSP have a Maxwellian velocity distribution, corresponding to a  mean space velocity of 130 km/s. This 
velocity distribution must be that of those binary systems that 
survived the first supernova explosion ~\cite{lyn1}. Therefore, we have supposed that 
binary systems which have also survived to the second explosion have a velocity distribution similar to 
those which have survived to the first explosion. 
The mean transverse velocity of observed MSP is around 90 km/s, which
corresponds exactly to the Maxwellian mean space velocity of 130 km/s
derived by Ref.~\refcite{cor} from parametric likelihood analysis techniques. 
So far, two binary systems have been detected through the second
pulsar: PSR B2303+46 and PSR J0737-3039 B. 
In order to observe two pulsars having 
millisecond like kick velocities, our simulations predict a total number of 
binaries $N_b = 730$, corresponding to a ratio  $N_p/N_b = 342$.
The advantage of our method is that we don't have to infer the
statistical properties of binary pulsars from the very small observed
sample and the only data we care about is the number of observed second born pulsars.
The derived ratio $N_p/N_b$ implies from Eq. (2), $f_b$ = 0.136 if the value 
$\beta_{ns}$ = 0.024 is adopted.
Notice that if the upper limit $f_b$ = 1 is imposed to Eq. (2), then we conclude 
that $\beta_{ns} \geq$ 0.0058 or that the 3-D velocity dispersion of the
kick distribution of binaries should satisfy $\sigma <$ 270 km/s,
e.g., lower than that of the single population.

For each simulated NS-NS binary, the separation, the orbital period
and the orbital eccentricity are available, allowing the calculation of the coalescence 
timescale $\tau$ due to emission of gravitational radiation.
The cumulative number of pairs with a coalescence timescale less than
$\tau$ varies as $N(<\tau) \propto lg(\tau)$, corresponding to a
probability distribution $P(\tau) = B/\tau$ (see Fig. 1) The
simulations indicate  a minimum coalescence timescale $\tau_0$ =
2$\times 10^5$ yr and a considerable number of systems having
coalescence timescales higher than the Hubble time. The normalized
probability $P(\tau)$ in the range $2\times 10^5$ up  to 20 Gyr
implies  B = 0.087.

\begin{figure}[pb]
\centerline{\psfig{file=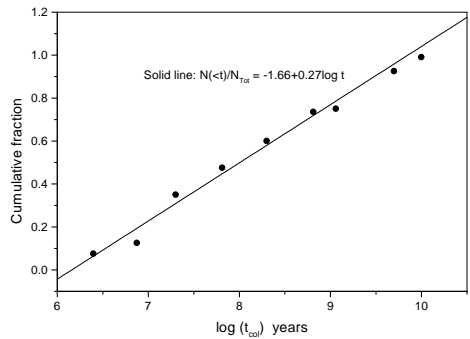,width=8cm}}
\vspace*{8pt}
\caption{The cumulative fraction of simulated NS-NS binary systems with coalescence time less or
equal $t$. Time is given is years and the solid curve shows the best fit to the simulated data. \label{f1}}
\end{figure}

The other important term required for the evaluation of the
coalescence rate from Eq.(1) is the star formation rate R$_*(t)$. In
most of population synthesis models, in the absence of a detailed star
formation theory, R$_*(t)$ is assumed to be proportional to the available mass of gas 
and under this (usual) assumption,  from the mass-balance 
equation, it results $R_*(t) \propto e^{-\alpha t}$. In fact, we will
use such an approach to model the evolution of an elliptical galaxy,
as we shall see in the next section. In some even more simplified galactic models the 
star formation rate is assumed to be constant.
However, for the Galaxy, the star formation history can presently be
derived from observations and most of recent studies have inferred
that the star formation activity is  non-monotonic with time
(Ref.~\refcite{rp} and references therein). Recent studies on the 
chromospheric activity index versus age relationship permitted the determination of 
ages for a sample of 552 stars from which was possible to reconstruct 
the star formation history of our Galaxy\cite{rp}. These data 
indicate enhanced periods of star formation at 1
Gyr, 2-5 Gyr and 7-9 Gyr ago, probably associated with accretion and
merger episodes  from which the disk grows and acquires angular
momentum\cite{pei}. Here we use
the results by Ref.~\refcite{rp} to compute numerically the
integral defined in Eq. (1).
Using the numbers obtained, it results for the present galactic NS-NS 
coalescence rate $\nu_S  = (1.7\pm 1.0)\times 10^{-5}$ yr$^{-1}$. The estimated error
is mostly due to uncertainties in the ratio $N_p/N_b$ derived from simulations.  
Similar population synthesis calculations to estimate the birth-rate
of compact binaries have also been performed by Ref.~\refcite{sch}. However some differences exist 
between their and the present calculations. In the one hand, Ref.~\refcite{sch} normalized their
birthrates to the type II+Ib/c supernova frequency, which was assumed
to be constant over the lifetime of the galactic disk. Moreover, they
have assumed that NS-NS binaries have the same natal kick distribution
as that observed for single pulsars. On the other hand, our
calculations took into account the star formation history of our
Galaxy and the normalization was derived from two simulation sets, which fixed respectively the 
fraction of bound NS-NS pairs and the ratio between binary
and single pulsars. However, it is worth mentioning that the cumulative number of
NS-NS pairs with a coalescence timescale less than $\tau$, resulting
from both set of simulations is similar.

\subsection{The coalescence rate in ellipticals}

Excluding the Milk Way,  no direct information is available for the
neutron star population and the star formation history of other
galaxies. Thus, the coalescence rate in extragalactic objects can only be estimated  
theoretically.  In this case, we have assumed for elliptical galaxies
the same values derived for the Milky Way for the parameters $f_b$, $\beta_{ns}$.
It is  worth mentioning that previous estimates of the present 
coalescent rate in ellipticals\cite{dfp} lead to
values which were about a half of that estimated for the
Galaxy. However, in that work the star formation efficiency was constrained essentially 
by the present amount of gas in ellipticals, a very uncertainty quantity. 
A grid of models for elliptical galaxies was recently built in
Ref.~\refcite{idi}, including the effects of mass-loss by a galactic wind, responsible for the enrichment of the
intra-cluster medium in metals. In these models, the parameters
characterizing the star formation efficency and the IMF were chosen by
an iterative procedure in order to reproduce the color-magnitude
diagram, (U-V) against M$_V$ of ellipticals in  Coma and Virgo
clusters. We take their model 3 as a representative of a typical E-galaxy, defined by an initial 
mass equal to 2$\times 10^{11}$ M$_{\odot}$ 
and a present luminosity L$_B$ = 2.9$\times 10^{10}$
L$_{B,\odot}$. The IMF in these models are slightly flatter than the Salpeter's law ($\gamma \approx$ 2.19) and 
the corresponding $\lambda$ parameter is
$\lambda$ = 8.71$\times 10^{-3}~M_{\odot}^{-1}$. It should be emphasized that
although the color-magnitude diagram imposes essentially constraints
on low mass stars, the models also explain quite well the relative abundance of 
the $\alpha$-elements, determined by massive stars, since they reproduce
the observed metallicity indices like Mg$_2$ and $<Fe>$~~\cite{idi}. Under these conditions, the estimated
present coalescence rate is $\nu_E$ = 8.6$\times 10^{-5}$
yr$^{-1}$. Notice that this value is about a factor of 5 {\it higher}
than the galactic rate estimated in the previous section. Thus, in spite of fact that today ellipticals 
have practically a non-existent (or a very low) star formation activity, since the bulk of their stars were formed
in the first 1-2 Gyr, the pairs merging  today  were copiously formed in the 
past with long coalescence timescales. A similar phenomenon occurs
with type Ia supernovae.  No type II supernovae are detected in ellipticals because these 
objects are the consequence of the evolution of massive stars, which have a rather short 
lifetime. However, type Ia  supernovae are
the only class detected in early type galaxies. This class of
supernova is the consequence of the evolution of binary systems constituted by 
intermediate mass progenitors, which evolve into white dwarfs. The explosion of the compact object 
depends on the distribution of the merging timescale and only  systems with long evolutionary 
timescales will explose by now.
\begin{figure}[pb]
\centerline{\psfig{file=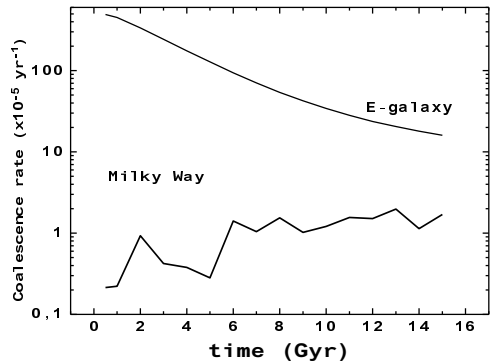,width=8cm}}
\vspace*{8pt}
\caption{Evolution of coalescence rates for the Milk Way and for a
  typical E galaxy. Notice that in the elliptical the bulk of stars is formed very early and the merging rate
reflects mainly the probability of a NS-NS binary to coalesce in a given timescale. For the
Galaxy, the star formation is continuous and intermittent, producing a modulation effect in
the coalescence rate history \label{f2}}
\end{figure}

In Fig. 2 we show the evolution of the merging rate for the Milky Way and for our representative
model of an elliptical galaxy. The smoothness observed in the curve
describing the evolution of the coalescence rate in the E galaxy is a
consequence of the fact that the bulk of the stars are quite old, being formed in a 
short time interval. Since the probability to have a merger is
inversely proportional to the time, the curve reflects essentially
this fact. This is not the case for our Galaxy, since its disk is
built gradually and the star formation rate may suddenly vary as new
matter is accreted.

\subsection{The local coalescence rate}

The models developed by Ref.~\refcite{idi} can equally reproduce the colors of S0 galaxies.
If we assume that the coalescence rate estimated above is typical for
E and S0 galaxies having absolute magnitudes of about M$_B$ = -20.7
and that the fraction of these objects is about 35\%, then the weighted local
coalescence rate can be written as
\begin{equation}
\nu_c = \nu_S(f_S + f_E\frac{\nu_E}{\nu_S}\frac{L_S}{L_E})
\end{equation}
where $f_S$ = 0.65 and $f_E$ = 0.35 are  respectively the adopted
fractions of spirals and (E+S0) galaxies, the $\nu_i$'s and $L_i$'s are the 
respective coalescence rates and luminosities.
>From this relation and values obtained before, the local mean weighted
coalescence rate is $\nu_c$ = 3.4$\times 10^{-5}$ yr$^{-1}$.
In the one hand, had we usually assumed the local merger rate equal to the galactic one, neglecting 
the contribution of (E+S0) galaxies, we
would have obtained  a value  smaller by a factor of two. On the other hand, the very 
recent galactic rate derived by Ref.~\refcite{kal3} seems to overestimate by a factor of 2.5 our 
results for the total local rate, which
include NS-NS mergers occurring in galaxies other than the Milky Way. 

\section{Expected detection rates}

Extrapolation from the local coalescence rate to the
expected rate within the volume $\frac{4\pi}{3}D^3$ is made
multiplying the local value by the factor $K_B(<D)$, which is defined as the ratio between 
the {\it total} blue luminosity within the 
considered volume and the Milky Way luminosity ($L_{B,MW} \approx
2.3\times 10^{10}\, L_{\odot}$). If galaxies were distributed
homogeneously in the local universe, the scale factor correcting the
local coalescence rate would simply be $K_B(<D) \approx
0.02D^3_{Mpc}$. This approximation is probably reasonable for
distances beyond 150 Mpc but underestimates the total  luminosity in
the B-band at lower distances, due to significant anisotropies in the
distribution of galaxies. Moreover, the sensitivity of interferometric detectors is 
 not isotropic, being larger to signals coming
from above and below the plane. In this paper,  the directional
sensitivity of the detector and the anisotropy in the distribution of
galaxies will not be taken into account as in Ref.~\refcite{nut}. However, real counts 
of galaxies inside a given volume will be 
considered, allowing a more realistic evaluation of the total luminosity.
Analyses of the spatial distribution of galaxies for $z < 0.033$ have
been performed by Ref.~\refcite{cou} (and references therein) using
the LEDA (Lyon-Meudon Extragalactic Database) catalog, which contains
over one million of galaxies covering  all-sky, including also a
sub-sample  of 134,000 objects having a measured redshift. The
completeness of the catalog was evaluated by counts of galaxies up to a given magnitude, excluding 
galaxies in the zone of avoidance (b$ < \mid 15^o\mid$), 
which includes superclusters like the Great Attractor. According to Ref.~\refcite{cou}, the 
redshift catalog is complete at level of
84\%  up to B = 14.5 and at a level of  90\% for galaxies brighter
than B  = 11.0. It is worth mentioning that up to B = 14.5, the total luminosity does not 
increase as the cube of the distance, but more slowly with
an exponent $\sim$ 2.5. Ref.~\refcite{cou} considered  different volume
limited samples, including only galaxies brighter than a given
absolute magnitude. We correct for the absence of faint galaxies, supposing a Schechter 
luminosity function in the magnitude interval  $-15 > M_B > -22$. 
Recent deep optical surveys in the zone of avoidance~\cite{kra,wou1,wou2} confirm the existence of a
large structure dubbed the  ''Great Attractor '' in the region $\mid
b\mid < 10^o$ and $316^o < l <  338^o$. This structure seems to be the intersection 
between the Centaurus Wall and the Norma supercluster, including
some other clusters as the Centaurus-Crux complex. The center of the
Great Attractor seems to be located at the Norma cluster (A3627) at
$V_z = 4844$ km/s~~\cite{kra} and up to now 4423 galaxies have been identified as members of 
this supercluster. The total mass
of such a complex is uncertain. A lower limit of  4$\times 10^{15}\,
M_{\odot}$ is derived from direct counts and the observed velocity
dispersion of the clusters, under the assumption of dynamical equilibrium. From large scale 
motions in this region, a mass as high as
5$\times 10^{16}\, M_{\odot}$ was obtained\cite{bur}. If we
assume a more conservative value of about $10^{16}\, M_{\odot}$ for
the total mass of the complex and a typical mass-to-luminosity ratio
of  150 $M_{\odot}/L_{B,\odot}$, then the Great Attractor alone would
give a relative contribution to the total luminosity of $K_B \approx$
2900, at a distance of 70 Mpc. This contribution was added  to the
total luminosity derived from data for distances D $\geq$ 70 Mpc. 
In Fig. 3, the calculated NS-NS coalescence rate ratio inside a volume
of radius D, given by local rate times the factor $K_B(<D)$, is shown
as a function of the distance D. Variations due to the Virgo cluster and the Great Attractor 
are indicated. Notice that rates corresponding to about
one event per year can be obtained only if the detector can probe
distances at least of the order of 90-100 Mpc.

\begin{figure}[pb]
\centerline{\psfig{file=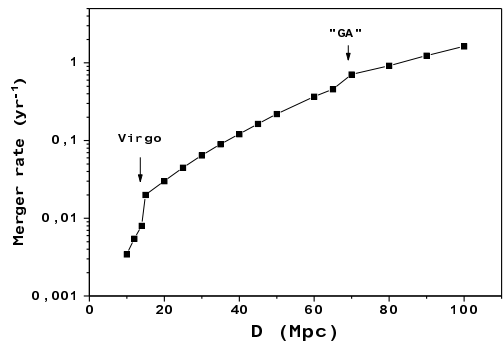,width=8cm}}
\vspace*{8pt}
\caption{Total expected NS-NS coalescence rate inside a volume of radius D as a
function of D \label{f3}}
\end{figure}

The strength of a signal observed in a given detector is characterized
by its signal-to-noise ratio $S/N$, which  measures the signal
amplitude in terms of the detector's noise characteristics. This
quantity depends on the source power spectrum, the noise spectral
density of the detector $S_n(\nu)$ in Hz$^{-1}$ and the adopted method
of data analysis. For merging NS-NS pairs, the optimum $S/N$ ratio is obtained by the so 
called  matched filter, namely,
\begin{equation}
(\frac{S}{N})^2 = 4\int_0^{\infty}\frac{\mid \tilde h(\nu)\mid^2}{S_n(\nu)}d\nu
\end{equation}
where $\mid\tilde h(\nu)\mid^2$ is the sum of  the square of the
Fourier transform of both polarization components. According to Ref.~\refcite{fin1}, the equation 
above can be written as
\begin{equation}
\frac{S}{N} =  8\Theta (\frac{r_0}{D})(\frac{{\cal M}}{1.2M_{\odot}})^{5/6}\zeta(\nu_{max})
\label{noiseratio}
\end{equation}
where $D$  is the distance-luminosity to the source and the parameter $r_0$ is given by the relation
\begin{equation}
r_0 = 9.25\times 10^{-22}\sqrt{I_{7/3}} \,\, Mpc
\end {equation}
with
\begin{equation}
I_{7/3} = (\frac{\nu_{\odot}}{\pi})^{1/3}\int_0^{\infty}\frac{d\nu}{\nu^{7/3}S_n(\nu)}
\end{equation}
where $\nu_{\odot}$ = 202.38 kHz. The term $\zeta(\nu_{max})$ is defined by
\begin{equation}
\zeta(\nu_{max}) = \frac{(\nu_{\odot}/\pi)^{1/3}}{I_{7/3}}\int_0^{2\nu_{max}}\frac{d\nu }{\nu^{7/3}S_n(\nu)}
\end{equation}
The spiral-in phase ends when the pair separation  is such that tidal
effects deform and may even disrupt the stars or when the last stable orbit is reached.
In Eq. (8),  $\nu_{max} \sim$ 750 Hz  corresponds to
the maximum orbital frequency attained nearly the inner most
circular orbit (see, for instance, Ref.~\refcite{fla}). The other
parameter appearing in Eq. (5) is the  ''chirp'' mass
 ${\cal M} = \mu^{3/5}M^{2/5}$, with $\mu$  and $M$ being respectively
 the reduced and the total mass of the system. The angular function
 $\Theta$ depends on geometrical projection factors of the detector,
 $F_+$ and $F_X$,  for  both polarization components (see, for
 instance, Ref.~\refcite{jar})  and on the inclination between the orbital angular 
momentum of the binary and the line of sight, namely,
\begin{equation}
\Theta^2 = 4\lbrack F_+^2(1 + cos^2i)^2 + 4F_X^2cos^2i\rbrack
\end{equation}
The probability distribution of $\Theta$ was discussed in Ref.~\refcite{fin2} and the authors have found 
an excellent approximation, which will be used here to estimate mean values
\begin{equation}
P(\Theta) = \frac{5\Theta(4 - \Theta)^3}{256} \,\,\, if  \,\,\, 0 \leq  \Theta \leq 4
\end{equation}. 
>From these equations, for a given $S/N$ ratio and detector
sensibility, the distance D probed by the instrument can be
evaluated. For VIRGO and LIGO we have used the noise spectral density
as  given respectively by Refs.~\refcite{pun1,dam} in a polynomial form, while
for advanced-LIGO we have adopted the expression given by
Ref.~\refcite{owe}. 
Under these conditions, the parameter $r_0$ for each detector is 7.6
Mpc for VIRGO, 8.0 Mpc for LIGO and 120 Mpc for advanced-LIGO. 
Adopting  S/N = 7.0, typical threshold for a false alarm rate of about
one per year, the maximum probed distances are 13, 14 and 207 Mpc for
VIRGO, LIGO and advanced-LIGO respectively. Using data from Fig. 2,
the mean expected event rates are one each 125 yr for LIGO, one each
148 yr for VIRGO and 6 events per yr for advanced-LIGO.
In a recent study, Refs.~\refcite{spa1,spa2}
have considered the spectral noise density of VIRGO in the context of chirp 
signal detection.
Four main spectral regions were identified, according to the dominant source
of noise: $\nu <$ 2 Hz (seismic noise);  2 Hz $< \nu <$ 52 Hz
(pendulum thermal noise); 52 Hz $< \nu <$ 148 Hz (mirror thermal
noise) and above 148 Hz (shot noise). Rather than directing efforts to
the lowest frequencies in the attempt to observe a large number of cycles, these 
analyses indicate that the reduction in the mirror thermal noise band
provides the highest gain in the S/N ratio. The direct dependence of
the mirror thermal noise power spectrum on the temperature suggests
the use of already existing cryogenic techniques. In parallel, also
the pendulum thermal and shot noises should be reduced in the mirror thermal noise band. If the 
total noise in such band 
could be reduced by a factor of 10, a gain by a factor of 8 can be
obtained, increasing the maximum probed distance up to 100 Mpc for S/N = 7 and allowing an 
expected detection rate of about 3 events 
every two years. A recently proposed  sound configuration\cite{pun2} for VIRGO would have a 
noise reduction through a large bandwidth and
a similar expected detection rate. In this configuration the pendulum  noise would reduced 
by a factor of 8, the mirror thermal noise by a factor of 7 and the shot noise by a factor of 4.
These results suggest that only the planned sensibility of the second
generation of interferometers will permit to probe the local universe
deep enough in order that at least few events should be seen in a monitoring period of one or two years.
An alternative strategy for the present generation of laser beam
interferometers is the search of these events by using a network of
detectors, since for a given false-alarm rate, the detection threshold
is lowered as the number of detectors increases\cite{schu,arn}. Data analysis of chirp signals detected by
a network of detectors, in particular the one constituted by the two 
LIGO (Hanford and Livingston) and VIRGO, was recently considered by Ref.~\refcite{pai}. They have 
estimated the increased detection
sensitivity for this particular combination of detectors and for a
false-alarm rate equal to one per year and a detection probability of
95\%. Using the numbers of their Table III, we derived a maximum
probed distance of 22 Mpc and an expected detection rate of about one event every 26 yrs.

\section{Conclusions}

In spite of the numerous investigations performed by different groups
in the past years on the coalescence rate of NS-NS binaries in the
Galaxy, the uncertainties are still quite large, due to several badly known aspects 
concerning the evolution of massive binaries.
In this work we have adopted a new approach to tackle this problem,
calculating the coalescence rate from the star formation history of the galaxy and from the 
probability per unit of time for a NS-NS system to merger in a timescale $\tau$.
In the case of the Galaxy, the star formation history was directly
derived from observations, using the cromospheric activity index as an age estimator. The data 
suggest that the star formation
rate is intermittent, consistent with a picture in which the disk is
gradually built by successive accretion episodes. Our numerical
simulations on the evolution of massive binaries show that only 2.4\%
of the pairs remain bound after the second supernova explosion. Our simulations indicate also
that the probability per unit of time for a NS-NS pair to coalesce in
a timescale $\tau$ is $P(\tau) = B/\tau$, in agreement with the early
estimate by Ref.~\refcite{dfp} and the recent simulations by Ref.~\refcite{sch}. Our approach 
indicate that the galactic
coalescence rate is presently $(1.7\pm 1.0)\times 10^{-5}\, yr^{-1}$,
a value which has remained more or  less constant in the past 9 Gyr. 
Elliptical galaxies also contribute to the local mean coalescence
rate. We have adopted as a typical E-galaxy a model from a grid able
to reproduce the observed color-magnitude diagram of early-type
galaxies in Virgo and Coma clusters. The star formation efficiency and
the initial mass function, required to evaluate the fraction of formed
massive stars, are fixed by fitting the photometric properties,
whereas the fraction of formed massive binaries and the fraction of
pairs which remain bound were taken to be equal to the galactic
values. Under these conditions, the typical coalescence rate for an
elliptical (or S0 galaxy) of absolute magnitude $M_B$ = -20.7 is about
$8.6\times 10^{-5}\, yr^{-1}$. Notice that this value is about a
factor of five higher than the galactic rate. 
The local coalescence rate, derived from an appropriate average
weighted by the relative luminosities and abundances, is $3.4\times 10^{-5}\, yr^{-1}$.
The expected coalescence within a volume of radius D was estimated by scaling
the local mean rate with respect to the ratio between the total
luminosity in the considered volume and the luminosity of the
Galaxy. Galaxy counts from the LEDA data base were used to estimate
the total luminosity. We have used the completeness estimates made by
Ref.~\refcite{cou} and the Schechter luminosity function to
correct for the absence of faint galaxies. The contribution of the
Great Attractor was included by assuming that the total mass of this
structure is about $10^{16}\, M_{\odot}$ and that the mean
mass-to-luminosity ratio is 150 $M_{\odot}/L_{B,\odot}$. Our results
indicate that a rate of about one event per year can only be obtained
if detectors can probe distances at least of the order of 80 
Mpc. This is not the case for the present generation of large laser
beam interferometers, since from their planned sensibility, one event
each 148 yr is expected for VIRGO whereas one event each 125 yr is
expected for LIGO. These rates are modified  when a network including
three detectors (Hanford, Livingston and VIRGO) are considered, since
in this case the expected rate corresponds to one event each 26 years.
Recently proposed improved VIRGO configurations may significantly
raise the expected detection rate up to 3 events every two years. For 
advanced-LIGO, 6 events per year are expected to be seen.

\section*{Acknowledgments}
T.R. is grateful to B.S Sathyaprakash for useful discussions and the authors
thanks the referee of this paper for his usefull comments. 
This research was partially supported by a PPARC (UK) post-doctoral
research associateship to T.R. 
A.S. was supported by the European
Space Agency (G. Colombo Senior Research Fellowship).

\end{document}